\documentstyle[aps,prl,twocolumn]{revtex}

\protect

\begin{document}
\draft
\title{Comment on ``Is The Nonlinear Meissner Effect Unobservable?''}
\author{Anand Bhattacharya$^{1}$, Igor \v {Z}uti\'{c}$^{2}$, Oriol T. Valls$^{1}$
and A.M. Goldman$^{1}$}
\address{$^1$School of Physics and Astronomy, University of Minnesota,\\
Minneapolis, MN 55455-0149}
\address{$^{2}$ Department of Physics, University of Maryland, \\
College Park, MD 20742-4111}
\date{\today}
\maketitle

\begin{abstract}
\end{abstract}




\narrowtext
In a recent Letter \cite{lxw} by Li, Hirschfeld and W\"{o}lfle (LHW) on
nonlocal effects in unconventional superconductors, it was suggested that
these effects might explain the null result for the nonlinear Meissner
effect (NLME) in our experiments \cite{buan,preprint} on optimally doped YBa$%
_{2}$Cu$_{3}$O$_{6.95}$ (YBCO) single crystals for which an appreciable
signal is predicted by theory \cite{ys,zv}.

We have no objection to the main part of the LHW letter, which deals with a
detailed calculation of the nonlocal effects\cite{kl}. However, the remarks
made about our experimental results do not directly follow from these
detailed calculations but are critically dependent on a qualitative argument
which fails to work for YBCO.

The qualitative argument relies on treating YBCO as a ``weakly 3-D'' system.
This leads LHW to the conclusion that nonlocal effects will wipe out the
NLME (for the geometry of our experiments) for fields below about $0.8\sim
1H_{c1}$. However, the estimate of $H_{c1}$ from the ``weakly 3-D'' argument
is\cite{lxw} $\Phi _{0}/(2\pi \lambda _{0}\lambda _{0c})$ (where $\Phi _{0}$
is the flux quantum, and $\lambda _{0}$ and $\lambda _{0c}$ are the{\it \ }
zero temperature penetration depths for currents flowing in the {\it a-b}
plane and along the {\it c-}axis respectively), which leads to a value of $%
H_{c1}$ of twenty Gauss or less. This is over an order of magnitude below
the experimental value of the field at which first flux penetration occurs
which is\cite{preprint} 300 Gauss or more\cite{buan}. In our experiments,
the samples used have typical dimensions of 1.5mm x 1.5mm x 50$\mu $m ({\it %
a }x {\it b} x {\it c}). The magnetic field is applied in the {\it a-b}
plane and given the sample geometry, most of the screening current flows in
the {\it a-b} plane with components along the nodal directions, with the
exception of the return currents that flow near the edges in the{\it \ c}%
-axis direction. Thus, we expect the field of first flux entry to be closer
to $\Phi _{0}/(2\pi \lambda _{0}^{2}),$ which is borne out by experiment.
For currents in the {\it a-b} plane, the nonlocal effects are very small,
much smaller than the NLME at fields of 300G. For the return currents, the
nonlocal effects may be relevant, {\it but} these do not contribute in any
way to NLME, as they have no components in the nodal directions. Thus, for
our experimental geometry, the nonlocal contributions are irrelevant.

The same conclusion can be reached even more starkly by starting from the
estimate of the characteristic nonlocal energy, given in LHW as $%
E_{nl}=\xi_{0c}\Delta_0/\lambda_0$. Quasiparticle effects will be
ineffective, due to nonlocality, for quasiparticles within an angle of less
than $\phi_{nl}$ from a node, where $\phi_{nl}$ is determined from the
condition $\Delta(\phi_{nl})/E_{nl}\sim 1$. This yields $\phi_{nl}\sim 0.001$
implying that the NLME requires an applied field $H>H_m$ with $H_m/H_0\sim
0.001$ where $H_0$ is\cite{zv} the characteristic field scale of the NLME.
Since $H_0$ is about\cite{zv,preprint} 8000 gauss, we have that $H_m$ is
about ten gauss, in rough agreement with the argument in the previous
paragraph.

To summarize: we find it quite plausible that the nonlocal effects indeed
render the NLME unobservable at fields below ten or twenty Gauss. Since the
experiments are performed at fields over one order of magnitude larger,
however, with the sample remaining in the Meissner state, it is obvious that
the explanation for our negative result must lie elsewhere. In our opinion%
\cite{preprint} the presence of at least a few percent component of
imaginary $s$ or $d_{xy}$ character in the gap remains the most likely
explanation.


%
%
%
%

\end{document}